\documentclass[12pt]{article}
\usepackage{array}
\usepackage{amsfonts}
\usepackage{amsmath}
\usepackage{amssymb}
\usepackage{bm}
\usepackage{chicago}
\usepackage{courier}
\usepackage{graphicx}
\usepackage{fullpage}
\usepackage{lineno}
\usepackage{verbatim}

\usepackage{color}

\providecommand{\U}[1]{\protect\rule{.1in}{.1in}}
\newcommand{\Remm}[1]{}

\newtheorem{model ass}[theo]{Model Assumptions}

\numberwithin{equation}{section}

\def\lboxit#1{\vbox{\hrule\hbox{\vrule\kern6pt
       \vbox{\kern6pt#1\kern6pt}\kern6pt\vrule}\hrule}}

\def\thick#1{\hbox{\rlap{$#1$}\kern0.25pt\rlap{$#1$}\kern0.25pt$#1$}}

\newfont{\fsc}{eusm10}                         

\modulolinenumbers[1]

\begin{document}

\title{Likelihood-free Bayesian inference for $\alpha$-stable models}
\author{G. W. Peters\footnote{Corresponding Author. Email: GarethPeters@unsw.edu.au}\:\,\footnote{School of Mathematics and Statistics, University of New South Wales, Sydney 2052, Australia.}, 
S.A. Sisson$^{\dagger}$ and Y. Fan$^{\dagger}$} 
\maketitle

\begin{abstract} 
$\alpha$-stable distributions are utilised as models for heavy-tailed noise  in many areas of statistics, finance and signal processing engineering. 
However, in general, neither univariate nor multivariate $\alpha$-stable models 
admit closed form densities which can be evaluated pointwise. This complicates the inferential procedure.
As a result,  $\alpha$-stable models are practically limited to the univariate setting under the Bayesian paradigm,
and to bivariate models  
under the classical framework.
In this article we develop a novel Bayesian approach to modelling univariate and multivariate $\alpha$-stable distributions based on recent advances in ``likelihood-free'' inference.
We present an 
evaluation of the performance of this procedure in 1, 2 and 3 dimensions, and provide an analysis of real daily currency exchange rate data. The proposed 
approach provides a feasible inferential methodology at a moderate computational cost.
\vspace{3mm}

\end{abstract}

\section{Introduction}

Models constructed with $\alpha$-stable distributions
possess several useful properties, including infinite variance,
skewness and heavy tails (\shortciteNP{zolotarev86};
\shortciteNP{alder+ft98}; \shortciteNP{samorodnitsky+g94};
\shortciteNP{nolan07}). $\alpha$-stable distributions provide no
general analytic expressions for the density, median, mode or
entropy, but are uniquely specified  by their
characteristic function, which has several
parameterizations. Considered as generalizations of the Gaussian
distribution, they are defined as the class of location-scale
distributions which are closed under convolutions. 
$\alpha$-stable distributions have found
application in many areas of statistics, finance and signal
processing engineering as models for impulsive, heavy tailed noise
processes (\shortciteNP{mandelbrot60}; \shortciteNP{fama65};
\shortciteNP{fama+r68}; \shortciteNP{nikias+s95}; \shortciteNP{godsill00}; \shortciteNP{melchiori06}).

The univariate $\alpha$-stable distribution is typically
specified by
 four parameters: $\alpha\in(0,2]$ determining the
rate of tail decay; $\beta\in\lbrack-1,1]$ determining the degree
and sign of asymmetry (skewness); $\gamma>0$ the scale (under some parameterizations); and
$\delta\in\mathbb{R}$ the location \shortcite{Levy24}. The parameter
$\alpha$ is termed the characteristic exponent, with small
and large $\alpha$ implying heavy and light tails respectively.
Gaussian ($\alpha=2, \beta=0$) and Cauchy
($\alpha=1,\beta=0$) distributions provide the only analytically tractable sub-members of this family. 
In general, as $\alpha$-stable models admit no closed form 
expression for the density which can be evaluated pointwise
 (excepting Gaussian and Cauchy 
 members), inference typically proceeds via the
characteristic function.

This paper is concerned with constructing both univariate and
multivariate Bayesian models in which the likelihood model is from
the class of $\alpha$-stable distributions. 
This is known to be a difficult problem. 
Existing methods for Bayesian $\alpha$-stable models are limited
to the univariate setting (\shortciteNP{buckle95}; \shortciteNP{godsill99};
\shortciteNP{godsill00}; \shortciteNP{lombardi07}; \shortciteNP{Casarin04};
\shortciteNP{Salas06}).

Inferential procedures for $\alpha$-stable models may be classified as auxiliary variable methods, inversion plus series expansion approaches and density estimation methods.
The auxiliary variable Gibbs sampler \shortcite{buckle95} increases the
dimension of the parameter space from $4$
($\alpha,\beta,\gamma$ and $\delta$) to $n+4$, where $n$ is the
number of observations. As strong correlations between parameters and large sample sizes are common in the $\alpha$-stable setting, this results in
a slowly mixing Markov chain since Gibbs moves are limited to moving parallel
to the axes (e.g. \shortciteNP{neal03}). 
Other Markov chain Monte Carlo (MCMC) samplers (\shortciteNP{dumouchel75,lombardi07}) 
adopt inversion techniques for
numerical integration of the characteristic function, employing
inverse Fourier transforms combined with a series expansion
\shortcite{bergstrom53} to accurately estimate distributional tails.
This is performed at each iteration of the Markov chain to evaluate the likelihood,
and is accordingly 
computationally intensive.
In addition the quality of the resulting approximation is sensitive to the spacing of the fast Fourier
transform grid and the point at which the series expansion begins (\shortciteNP{lombardi07}).

Univariate density estimation methods include integral representations
\shortcite{zolotarev86},  
parametric mixtures (\shortciteNP{Nolan97};
\shortciteNP{McCulloch98})
and numerical estimation through splines and series expansions
(\shortciteNP{Nolan01};
\shortciteNP{Nolan08}). \shortciteN{McCulloch98} approximates symmetric
stable distributions using a mixture of Gaussian and
Cauchy densities. \shortciteN{doganoglu+m98}
 and \shortciteN{Mittnik91} approximate the $\alpha$-stable density
through spline polynomials, and \shortciteN{kuruoglu+mgf97} via a mixture of
Gaussian distributions. 
Parameter estimation has  been performed by an expectation-maximization (EM)
algorithm \shortcite{Lombardi06}  and by
method of (partial) moments
(\citeNP{Press72,Weron05}).
Implemented within an MCMC sampler, such density estimation methods would be highly computational. 

None of the above methods easily generalize to the multivariate setting.
It is currently only practical to numerically evaluate two
dimensional $\alpha$-stable densities via inversion of the characteristic function. Here the required computation
is a
function of $\alpha$ and the number and spread of masses in the
discrete spectral representation (\shortciteNP{Nolan01};
\shortciteNP{Nolan97}). Beyond two dimensions this procedure becomes untenably slow with limited accuracy.

In this article we develop practical Bayesian inferential methods to fit univariate and multivariate $\alpha$-stable models.
To the best of our knowledge,
no practical Bayesian methods have been developed for the multivariate model 
as the required computational complexity 
increases dramatically with model dimension.
The same is true of classical methods beyond two dimensions.
Our approach is based on recent developments in ``likelihood-free'' inference,
which permits approximate posterior simulation for Bayesian models  without the need to explicitly evaluate
the likelihood. 

In Section \ref{sec:abc} we briefly introduce likelihood-free
inference and the sampling framework used in this article. Section \ref{sec:alpha-stable} presents
the Bayesian $\alpha$-stable model, with a particular
focus
on summary 
statistic specification, a critical component of likelihood-free inference.
We provide an evaluation of the performance of the proposed methodology in  Section \ref{sec:examples},
based on controlled simulation studies in 1, 2 and 3 dimensions.
Finally, in Section \ref{sec:realdata} we demonstrate an analysis of real daily currency data under
both univariate and multivariate settings, and 
provide comparisons with existing methods.
We conclude with a discussion.

\section{Likelihood-free models}
\label{sec:abc}

Computational procedures to simulate from posterior distributions, $\pi(\theta|y)\propto\pi(y|\theta)\pi(\theta)$, of parameters $\theta\in\Theta$ given observed data $y\in\mathcal{X}$, are well established (e.g. \shortciteNP{brooks10}).
However when pointwise evaluation of the likelihood function $\pi(y|\theta)$ is computationally prohibitive or intractable, alternative procedures are required.
Likelihood-free methods (also known as {\it approximate Bayesian computation}) permit simulation from an approximate posterior model while circumventing explicit evaluation of the likelihood function
(\shortciteNP{tavare+bgd97}; \shortciteNP{beaumont+zb02};  \shortciteNP{Marjoram03}; \shortciteNP{sisson+ft07}; \shortciteNP{ratmann+ahwr09}).

Assuming data simulation  $x\sim\pi(x|\theta)$ under the model given $\theta$ is easily obtainable,
likelihood-free methods embed the posterior $\pi(y|\theta)$ within an
augmented model
\begin{equation}
\label{eqn:joint-posterior}
    \pi_{LF}(\theta,x|y)\propto\pi_\epsilon(y|x,\theta)\pi(x|\theta)\pi(\theta),
\end{equation}
where $x\sim\pi(x|\theta)$, $x\in{\mathcal X}$, is an auxiliary parameter on the
same space as the observed data $y$.
The function
$\pi_\epsilon(y|x,\theta)$ is typically a standard smoothing kernel (e.g. \citeNP{blum09}) with scale parameter $\epsilon$, which  weights 
the intractable posterior
with high values in regions when the 
observed data $y$ and auxiliary data $x$ are similar.
For example, uniform kernels are commonplace in likelihood-free models (e.g. \shortciteNP{Marjoram03,sisson+ft07}), although alternatives such as Epanechnikov \shortcite{beaumont+zb02} and Gaussian kernels \shortcite{peters+fs08} provide improved efficiency.
The resulting approximation to the true posterior
target distribution 
\begin{equation}
\label{eqn:marginal-posterior}
    \pi_{LF}(\theta|y)\propto\int_{\mathcal X}\pi_\epsilon(y|x,\theta)\pi(x|\theta)\pi(\theta)dx
    =
    \pi(\theta)\mathbb{E}_{\pi(x|\theta)}[\pi_\epsilon(y|x,\theta)]
\end{equation}
improves as $\epsilon$ decreases, and exactly recovers the target posterior as $\epsilon\rightarrow 0$, as then $\lim_{\epsilon\rightarrow 0}\pi_\epsilon(y|x,\theta)$ becomes a point mass at $y=x$ \shortcite{Reeves05}.

Posterior simulation from $\pi_{LF}(\theta|y)$ can then proceed via standard simulation algorithms, replacing pointwise evaluations of $\pi_{LF}(\theta|y)$ with Monte Carlo estimates through the expectation (\ref{eqn:marginal-posterior}), based on draws $x^1,\ldots,x^P\sim\pi(x|\theta)$ from the model (e.g. \shortciteNP{Marjoram03}). Alternatively, simulation from the joint posterior $\pi_{LF}(\theta,x|y)$ is available by contriving to cancel the intractable likelihoods $\pi(x|\theta)$ in sample weights or acceptance probabilities. For example, importance sampling from the prior predictive distribution $\pi(\theta,x)=\pi(x|\theta)\pi(\theta)$ results in an importance weight of $\pi_{LF}(\theta,x|y)/\pi(\theta,x)\propto\pi_\epsilon(y|x,\theta)$, which is free of likelihood terms. 
See \shortciteN{sisson+pfb08} for a discussion of marginal and joint-space likelihood-free samplers.

In general, the distribution of $\pi(x|\theta)$ will be diffuse, unless $x$ is discrete and $\dim(x)$ is small. Hence, generating $x\sim\pi(\cdot|\theta)$ with $x\approx y$ is improbable for realistic datasets $y$, and as a result the degree of computation required for a good likelihood-free approximation $\pi_{LF}(\theta|y)\approx\pi(\theta|y)$ (i.e. with small $\epsilon$) will be prohibitive.
In practice, the function $\pi_\epsilon(y|x,\theta)$ is expressed through low dimensional vectors of summary statistics, $S(\cdot)$, such that $\pi_\epsilon(y|x,\theta)$ weights the intractable posterior through (\ref{eqn:joint-posterior}) with high values in regions where $S(y)\approx S(x)$. 

If $S(\cdot)$ is sufficient for $\theta$, then letting $\epsilon\rightarrow 0$ recovers $\pi_{LF}(\theta|y)=\pi(\theta|y)$ as before, but with more acceptable computational overheads, as $\dim(S(x))<<\dim(x)$.
As sufficient summary statistics are generally unavailable,  the use of non-sufficient statistics is commonplace. The effect of less efficient estimators of $\theta$ in (\ref{eqn:marginal-posterior}) is a more diffuse approximation of $\pi(\theta|y)$. Hence the choice of summary statistics in any application is critical, with the ideal being low-dimensional, efficient and near-sufficient.

In this article, we implement the likelihood-free sequential Monte Carlo sampler of \shortciteN{peters+fs08}, detailed in Appendix A.
As the class of particle-based algorithms is the most efficient currently available in likelihood-free computation (e.g. \shortciteNP{mckinley+cd09}), and within this class, the sampler of \shortciteN{peters+fs08} is the only one to allow non-uniform  functions $\pi_\epsilon(y|x,\theta)$, 
this sampler provides the best combination of efficient simulation and flexible modelling.

\section{Bayesian $\alpha$-stable models}
\label{sec:alpha-stable}

We now develop univariate and multivariate Bayesian
$\alpha$-stable models.
Unlike existing methods,
likelihood-free inference is independent of model parameterization.

\subsection{Univariate $\alpha$-stable Models}
\label{sec:uni}

Denote the characteristic function of $n$ i.i.d. univariate $\alpha$-stable
distributed random variables $X_1,\ldots,X_n$ by $\Phi_{X}(t).$
A popular and convenient  parameterization is
\begin{equation}\label{AlphastableCharacteristicFn}
\Phi_{X}\left(  t\right)  =\left\{
\begin{array}{ll}
\exp\left(  i\delta t-\gamma^{\alpha}\left\vert t\right\vert
^{\alpha}\left[ 1+i\beta\tan\frac{\pi\alpha}{2}\text{sgn}\left(
t\right) \left( \left\vert \gamma t\right\vert
^{1-\alpha}-1\right) \right] \right) & \text{ \ if }\alpha
\neq1\\
\exp\left(  i\delta t-\gamma\left\vert t\right\vert \left[
1+i\beta\frac {2}{\pi}\text{sgn}\left(  t\right)  \ln\left(
\gamma\left\vert t\right\vert \right) \right]  \right) & \text{ \
if }\alpha=1,
\end{array}
\right.
\end{equation}
where $\text{sgn}\left(t\right) =\frac{t}{\left\vert t\right\vert
}$ and $i^2=-1$ (e.g. \citeNP{samorodnitsky+g94}).
Many alternative parameterizations are detailed in \shortciteN{nolan07} and \shortciteN{zolotarev86}.
Under (\ref{AlphastableCharacteristicFn}), the intractable stable density 
function is continuous and
unimodal, taking support on
$\left(
-\infty,0\right) $ if $\alpha<1,\beta=-1$; $\left(
0,\infty\right) $ if $\alpha<1,\beta=1$ and $\left(
-\infty,\infty\right)  $ otherwise. 

Efficient simulation of auxiliary data, $x\sim\pi(x|\theta)$, under the model is critical for the performance of likelihood-free methods (Section \ref{sec:abc}). Here, it is straightforward to generate $\alpha$-stable variates under the model defined by the characteristic function (\ref{AlphastableCharacteristicFn})
(e.g. 
\shortciteNP{devroye86}; \shortciteNP{nolan07}). This approach is provided in Appendix B.

\subsubsection{Summary statistics}
\label{sec:uni-ss}

A key component of likelihood-free inference is the availability of low-dimensional, efficient and near-sufficient
summary statistics.
Since $\alpha$-stable models can possess
infinite variance ($\alpha>1$) and infinite mean ($\alpha<1$), this choice
must be made with care.
Here we present several candidate summary vectors, $S_1$--$S_5$, previously utilized
for parameter estimation in the univariate $\alpha$-stable model. 
In Section \ref{sec:examples} we evaluate the
performance of these vectors, and provide informed recommendations for the choice of summary statistics under the likelihood-free framework.

\vskip 0.5cm
\noindent \bm{$S_1$}\textit{ McCulloch's Quantiles}\\
\shortciteN{McCulloch86} and \shortciteN{McCulloch98} estimate model parameters based on sample quantiles, while correcting for 
estimator
skewness due to the evaluation of $\widehat{q}_{p}(x)$,  the $p^{th}$  quantile of $x$, with a finite
sample. 
Here, the data $x_{(i)}$ are arranged in
ascending order and matched with $\widehat{q}_{s(i)}(x)$, where $s\left(  i\right)  =\frac
{2i-1}{2n}$. Linear interpolation to $p$ from the two
adjacent $s(i)$ values then establishes $\widehat{q}_{p}(x)$ as a consistent estimator of the true quantiles. 
Inversion of the functions
\[
\widehat{v}_{\alpha}=\frac{\widehat{q}_{0.95}(\cdot)
-\widehat {q}_{0.05}(\cdot)}{\widehat{q}_{0.75}(\cdot)-\widehat {q}_{0.25}(\cdot)},\:
\widehat{v}_{\beta}=\frac{\widehat {q}_{0.95}(\cdot)
+\widehat{q}_{0.05}(\cdot)-2\widehat {q}_{0.5}(\cdot)}{\widehat{q}_{0.95}(\cdot)-\widehat
{q}_{0.05}(\cdot)},\:\widehat{v}_{\gamma}=\frac{\widehat
{q}_{0.75}(\cdot)-\widehat{q}_{0.25}(\cdot)}{\gamma}
\]
then provides estimates of 
$\alpha, \beta$ and $\gamma$. Note that
from a computational perspective,
inversion of $v_{\alpha}, v_{\beta}$ or $v_{\gamma}$ is not required under
likelihood-free methods.
Finally, we estimate $\delta$ by the sample mean $\bar{X}$ (when $\alpha > 1$). 
Hence 
$S_1(x)  
=(\widehat{v}_{\alpha},\widehat{v}_{\beta},\widehat{v}_{\gamma},\bar{x}).$

\vskip 0.5cm 
\noindent \bm{$S_2$} \textit{ Zolotarev's Transformation} \\
Based on a transformation of data from the $\alpha$-stable family  $X\rightarrow Z$,
\shortciteN{zolotarev86} (p.16) provides
an alternative parameterization of the $\alpha$-stable model  
$(\alpha,\beta,\gamma,\delta)\leftrightarrow(\nu,\eta,\tau)$
with a characteristic function of the form
\begin{equation}
\label{ZoltorevZparam}
\log\Phi
_{Z}\left(  t\right)  =-\exp\left\{
\nu^{-\frac{1}{2}}\left[\log\left\vert t\right\vert
+\tau-i\frac{\pi}{2}\eta\, \text{sgn}\left(  t\right) \right]
+\mathbb{C}\left(  \nu^{-\frac{1}{2}}-1\right)  \right\},
\end{equation}
where $\mathbb{C}$ is Euler's constant, and where
$\nu\geqslant\frac{1}{4}$, $\left\vert \eta \right\vert
\leq\min\{1,2\sqrt{\nu}-1\}$ and $\left\vert
\tau\right\vert <\infty.$ 
This parameterization
has the advantage that 
logarithmic moments have simple
expressions in terms of parameters to be estimated.
For a fixed constant $0<\xi\leq\frac{1}{2}$ (\shortciteNP{zolotarev86} recommends $\xi=0.25$) 
and for integer $n/3$,
the transformation is 
\[
Z_{j}=X_{3j-2}-\xi X_{3j-1}-(1-\xi) X_{3j}, \qquad
j=1,2,\ldots,n/3.
\]
Defining $
V_{j}=\log\left\vert Z_{j}\right\vert$ and $U_{j}=\text{sgn}(X_{j})$, estimates for $\nu,\eta$ and $\tau$ are then
given by
\[
\widehat{\nu}=\max\{  \tilde{\nu},\left(
1+\left\vert \widehat{\eta}\right\vert \right)  ^{2}/4\},
\quad\widehat{\eta}=\mathbb{E}\left(
U\right), \quad\widehat{\tau}=\mathbb{E}\left(  V\right),
\]
where $\tilde{\nu}=\frac{6}{\pi^{2}}S^2(V)-\frac{3}{2}S^2(U)+1$, using sample variances $S^2(V)$ and $S^2(U)$.
As before, $\delta$ is estimated by $\bar{X}$ (for $\alpha>1$), and so $S_2(x)=(
\widehat{\nu},\widehat{\eta},\widehat{\tau},\bar{x})$.

\vskip 0.5cm
\noindent \bm{$S_3$}\textit{ Press's Method Of Moments} \\
For $\alpha\neq1$ and unique evaluation points $t_{1}, t_{2}, t_{3}, t_{4}$,  the method of moments equations obtained from
$\log\Phi_X(t)$
can be solved to obtain
(\shortciteNP{Press72}; \shortciteNP{Weron05})
\begin{eqnarray*}
\log(\widehat{\gamma})   &  =& \frac{\log\left\vert t_{1}
\right\vert \log(-\log\left\vert \Phi(  t_{2})
\right\vert)  -\log\left\vert t_{2}\right\vert \log(
-\log\left\vert
\Phi( t_{1})  \right\vert) }{\log\left\vert t_1/t_2\right\vert },
\quad
\widehat{\alpha}=\frac{\log\frac
{\log\left\vert \Phi\left(  t_{1}\right)  \right\vert
}{\log\left\vert
\Phi\left(  t_{2}\right)  \right\vert }}{\log\left\vert t_1/t_2\right\vert }\\
\left.  
\widehat{\beta}\right\vert
_{\widehat{\alpha},\widehat{\gamma}}  & = &
\frac{\widehat{u}\left(  t_{4}\right)/t_{4}-
\widehat{u}\left(  t_{3}\right)/t_{3}}{(  \left\vert t_{4}%
\right\vert ^{\widehat{\alpha}-1}-\left\vert t_{3}\right\vert
^{\widehat{\alpha}-1})
\widehat{\gamma}^{\widehat{\alpha}}\tan\left(  \frac{\widehat{\alpha}\pi}%
{2}\right)  },
\quad 
\left. \widehat{\delta}\right\vert _{\widehat{\alpha}} =
\frac{\left\vert t_{4}\right\vert
^{\widehat{\alpha}-1}\widehat {u}\left(  t_{3}\right)
/t_{3}-\left\vert t_{3}\right\vert ^{\widehat
{\alpha}-1}\widehat{u}\left(  t_{4}\right)
/t_{4}}{\left\vert t_{4}\right\vert
^{\widehat{\alpha}-1}-\left\vert t_{3}\right\vert
^{\widehat{\alpha}-1}}
\end{eqnarray*}
where $\widehat{u}(t)  = \tan^{-1}[\sum_{i=1}^n\cos(tx_i)/\sum_{i=1}^n\sin(tx_i)]$.
We adopt the evaluation points $t_1=0.2, t_2=0.8, t_3=0.1$ and $t_4=0.4$ as recommended by \shortciteN{Koutrouvelis80}, and accordingly obtain 
$S_3(x)=(\widehat{\alpha},\widehat{\beta
},\widehat{\gamma},\widehat{\delta}).$

\vskip 0.5cm
\noindent \bm{$S_4$}\textit{ Empirical Characteristic Function} \\
The empirical characteristic function,
$
	\widehat{\Phi}_X(t)  =\frac{1}{n}\sum_{j=1}^ne^{itX_j}
$
for $t \in (-\infty,\infty)$, can be used as the basis for summary statistics
when standard statistics are not available. E.g. this may occur through the non-existence of moment generating functions.
Hence, we specify
$S_4(x)=(\widehat{\Phi}_X(t_1),\ldots,\widehat{\Phi}_X(t_{20}))$ where $t_i\in\{\pm0.5,\pm 1,\pm 1.5,\ldots,\pm5\}$.

\vskip 0.5cm
\noindent \bm{$S_5$}{\textit{ Mean, Quantiles and Kolmogorov-Smirnov Statistic} \\
The Kolmogorov-Smirnov statistic is defined as
$ KS(X)= \sup_z |F^X_n(z)-F^Y_n(z)|$, the largest absolute deviation between the empirical cumulative distribution functions of auxiliary ($X$) and observed ($Y$) data,
where $F^X_n(z)={1 \over n}\sum_{i=1}^n I_{(X_i\leq z)}$
and
$I_{(X_i\leq z)}=1$
if $X_i \leq z$ and 0 otherwise.
We specify $S_5(x)=(\bar{x},\{\widehat{q}_{p}(x)\},KS(x))$, where the set of sample quantiles $\{\widehat{q}_{p}(x)\}$ is determined by 
$p\in\{0.01,0.05,0.1,0.15,\ldots,0.9,0.95,0.99\}$.

\vskip 0.5cm
Likelihood-free inference may be implemented under any parameterization which permits data generation under the model, and for which the summary statistics are well defined. From the above $S_1$--$S_5$ are jointly well defined for $\alpha>1$. 
Hence, to complete the specification of the univariate $\alpha$-stable model 
we 
adopt the independent uniform priors 
$\alpha\sim\text{U}[1.1,2]$, $\beta\sim\text{U}[-1,1]$,
$\gamma\sim\text{U}[0,300]$ and $\delta\sim\text{U}[-300,300]$ (e.g. \shortciteNP{buckle95}).
Note that 
the prior for $\alpha$ has a restricted domain, reflecting the use of sample moments in $S_1$--$S_3$ and $S_5$. For $S_4$ we may adopt 
$\alpha\sim\text{U}(0,2]$.

\subsection{Multivariate $\alpha$-stable Models}
\label{sec:mvas}

Bayesian 
model specification and simulation in the
multivariate $\alpha $-stable setting is 
challenging
(\shortciteNP{Nolan01}; \shortciteNP{Nolan08}; \shortciteNP{samorodnitsky+g94}).
Here we follow 
\shortciteN{Nolan08}, who defines the
multivariate model for the random vector $\bm{X}=(X_1,\ldots,X_d)\in\mathbb{R}^d$ through the
 functional equations
\begin{eqnarray*}\label{eqn:functionals_multi}
\sigma^{\alpha}\left(
\mathbf{t}\right)   & =&\int_{S_{d}}\left\vert
\left\langle \mathbf{t},\mathbf{s}\right\rangle \right\vert ^{\alpha}%
\Gamma\left(  d\mathbf{s}\right); 
\quad \beta\left(  \mathbf{t}\right)  =\sigma^{-\alpha}\left(  \mathbf{t}%
\right)  \int_{S_{d}}\text{sgn}\left(  \left\langle \mathbf{t},\mathbf{s}%
\right\rangle \right)  \left\vert \left\langle \mathbf{t},\mathbf{s}%
\right\rangle \right\vert ^{\alpha}\Gamma\left(  d\mathbf{s}\right) \\
\mu\left(  \mathbf{t}\right)   &  =&\left\{
\begin{array}
{ll}
\left\langle \mathbf{t},\boldsymbol{\mu}^{0}\right\rangle & \text{ \ \ }\alpha\neq1\\
\left\langle \mathbf{t},\boldsymbol{\mu}^{0}\right\rangle
-\frac{2}{\pi}\int_{S_d}\left\langle
\mathbf{t},\mathbf{s}\right\rangle \ln\left\vert \left\langle \mathbf{t}%
,\mathbf{s}\right\rangle \right\vert \Gamma\left(
d\mathbf{s}\right) & \text{ \ \ }\alpha=1
\end{array}
\right.
\end{eqnarray*}
where 
$\bm{t}=(t_1,\ldots,t_d)$, $\bm{s}=(s_1,\ldots,s_d)$,
$\left\langle
\bm{t},\bm{s}\right\rangle = \sum_{i=1}^dt_is_i$,
$\mathbb{S}^{d}$ denotes the unit $d$-sphere, $\Gamma\left( d\mathbf{s}\right)
$ denotes the unique spectral measure, and
$\sigma^{\alpha}\left( \mathbf{t}\right) $ represents scale,
$\beta\left( \mathbf{t}\right)  $ skewness and $\mu\left(
\mathbf{t}\right)$ location (through the vector  $\boldsymbol{\mu}^0=(\mu_1^0,\ldots,\mu_d^0)$). 
Scaling properties of the functional equations, e.g. $\mu(r\mathbf{t})=r\mu(\mathbf{t})$, mean that it is sufficient to consider them on the unit sphere 
\cite{Nolan97}. 
The corresponding 
characteristic function is 
\[
\Phi_{\mathbf{X}}\left(  \mathbf{t}\right)  =\mathbb{E}\exp\left(
i\left\langle \mathbf{X},\mathbf{t}\right\rangle \right)  =\exp\left(
-I_{\mathbf{X}}\left(  \mathbf{t}\right)  +i\left\langle \boldsymbol{\mu}
^{0},\mathbf{t}\right\rangle \right)
\]
with
$
I_{\mathbf{X}}(\mathbf{t}) ={\textstyle\int_{\mathbb{S}^{d}}}
\psi_{\alpha}\left(  \left\langle
\mathbf{t},\mathbf{s}\right\rangle \right) \Gamma(
d\mathbf{s}),
$
where the function $\psi_{\alpha}$ is given by
\[
\psi_{\alpha}\left(  u\right)  =\left\{
\begin{array}
{ll} \left\vert u\right\vert ^{\alpha}\left(  1-i\text{sgn}\left(
u\right)  \tan\left(
\frac{\pi\alpha}{2}\right)  \right) & \text{ \ \ }\alpha\neq1\\
\left\vert u\right\vert ^{\alpha}\left(
1-i\frac{2}{\pi}\text{sgn}\left( u\right) \ln\left\vert
u\right\vert \right) & \text{ \ \ }\alpha=1.
\end{array}
\right.
\]
The spectral measure $\Gamma(\cdot)$ and location vector $\boldsymbol{\mu}^0$ uniquely
characterize the multivariate distribution
(\shortciteNP{samorodnitsky+g94}) and carry essential information
relating to the dependence between the elements of $\mathbf{X}$.
The continuous spectral measure
is typically well approximated by a discrete set of $k$ Dirac masses
$
\Gamma\left(  \cdot\right)  =\sum\limits_{j=1}^{k}w_{j}\delta_{\mathbf{s}_{j}
}(\cdot)
$
(e.g. \shortciteNP{byczkowski93})
where $w_{j}$ and $\delta_{\mathbf{s}_{j}}(\cdot)$
respectively denote the weight and Dirac mass of the $j^{th}$ spectral
mass at location
$\mathbf{s}_{j}\in \mathbb{S}^{d}$. 
By 
simplifying
the 
integral 
in $I_{\mathbf{X}}(\mathbf{t})$, 
computation with the characteristic function 
$
	\Phi_{\mathbf{X}}^{\ast}(\mathbf{t})  =\exp\{
	-\sum_{j=1}^{k}  w_{j}\psi_{\alpha}( \left\langle \mathbf{t},\mathbf{s}_{j}\right\rangle)\}
$
becomes tractable
and data generation from the 
distribution 
defined by
$\Phi_{\mathbf{X}}^{\ast}(\bm{t})$ 
is efficient (Appendix B).
As with the univariate case (\ref{AlphastableCharacteristicFn}), standard parameterizations of
$\Phi_{\mathbf{X}}^{\ast}\left(  \mathbf{t}\right)$ will be
discontinuous at $\alpha=1$, 
resulting in poor
estimates of location and
$\Gamma(\cdot)$.
In the multivariate setting this is
overcome by 
Zolotarev's M-parameterization (\shortciteNP{Nolan01}; \shortciteNP{Nolan08}).
 Although likelihood-free methods are parameterization independent,
it is sensible
to work with  models with good likelihood properties.

In a Bayesian framework we 
parameterize the model via the spectral mass,
which involves estimation of
the weights $\boldsymbol{w}=(w_{1},\ldots,w_{k})$ and  locations $\mathbf{s}_{1:k}
\in\mathbb{S}^{d\times k}$
of $\Gamma(\cdot)$.
 For $k=2$ this corresponds
to $\mathbf{s}_{i}=\left( \cos\phi_{i},\sin\phi_{i}\right) \in
\mathbb{S}^{2}$.
More generally, for $\mathbf{s}_{i}= \left(
s_{i}^{1},\ldots,s_{i}^{d}\right) \in \mathbb{S}^{d}$, we use
hyperspherical coordinates $\boldsymbol{\phi}_i=(\phi^i_{1},\ldots,\phi^i_{d-1})$, where
\[
\phi_{d-1}^{i} = \tan^{-1}\left(
\frac{s^{d}_{i}}{s^{d-1}_{i}}\right),
\quad \ldots\quad,\quad \phi_{1}^{i} = \tan^{-1}\left(
\frac{\sqrt{\left( s^{d}_{i}\right)
^{2}+\left(  s^{d-1}_{i}\right)^{2}+\ldots+\left( s^{2}_{i}\right)^{2}}%
}{s^{1}_{i}}\right).
\]
We define priors for the parameters $\left(
\boldsymbol{w},\boldsymbol{\phi}_{1:k},\bm{\mu}^{0},\alpha\right)$, with $\boldsymbol{\phi}_{1:k}=(\boldsymbol{\phi}_1,\ldots,\boldsymbol{\phi}_k)$, 
as $\boldsymbol{w}\sim
\text{Dirichlet}(s,\ldots,s)$, $\phi_{j'}^{i}\sim
\text{U}(0,2\pi)$, $\mu_{j}^{0}\sim
\text{N}(\xi,\kappa^{-1})$, $\alpha\sim
\text{U}(0,2)$ for $i=1,\ldots,k$, $j=1,\dots,d$, $j'=1,\ldots,d-1$
and
impose the ordering constraint $s_{i-1}^{1} \leq s_i^1$ for all $i$ on the first element of each vector
$\bm{s}_i$. 
Note that
by treating the weights and locations of the spectral masses
as unknown parameters, 
 these may be identified
with those regions of the spectral measure 
with significant posterior mass.
This differs with
the approach of \shortciteN{Nolan97} where the spectral mass is
evaluated at a large number of deterministic grid locations. 
In estimating $\Gamma(\cdot)\mid \mathbf{X}$
a significant reduction in the number
of required projection vectors is achieved (Sections \ref{sec:mv-summary} and \ref{sec:ss;mv}).
Further, the above prior specification does not penalize placement of spectral masses in close proximity.
While this proved adequate for the presented analyses, alternative priors may usefully inhibit spectral masses at similar locations
(\shortciteNP{Piev&G1998}).

\subsubsection{Summary statistics}
\label{sec:mv-summary}

\noindent \bm{$S_6$}{ \textit{ Nolan, Panorska \& McCulloch Projection Method}\\
For the $d$-variate $\alpha$-stable observations, $\mathbf{X}_i, i=1,\ldots,n$, 
we take projections of $\mathbf{X}_i$ onto a unit hypersphere in the direction $\mathbf{t}\in\mathbb{S}^d$. 
This produces a set of $n$ univariate values, $\mathbf{X}^{\mathbf{t}}=(X^{\mathbf{t}}_1,\ldots,X^{\mathbf{t}}_n)$, where $X_i^{\mathbf{t}}=\left\langle
\mathbf{X}_{i},\mathbf{t}\right\rangle $. 
The information in $\mathbf{X}^{\mathbf{t}}$ can then be summarized by any of the univariate summary statistics $S_1(\mathbf{X}^{\mathbf{t}}),\ldots,S_5(\mathbf{X}^{\mathbf{t}})$.
This process is repeated for multiple projections over $\mathbf{t}_1,\ldots,\mathbf{t}_\tau$.
With the location parameter $\bm{\mu^0}$
estimated 
by $\bar{\mathbf{x}}$, for sufficient numbers of projection vectors, $\tau$,  the summary statistics $S_6(\mathbf{X})=(\bar{\mathbf{x}}, S_s(\mathbf{X}^{\mathbf{t}_1}),\ldots,S_s(\mathbf{X}^{\mathbf{t}_\tau}))$ for some $s\in\{1,2,3,4,5\}$, will capture much of the information contained in the
multivariate data, if $S_s$ is itself informative.
The best choice of univariate summary vector $S_s$ will be determined in Section
\ref{sec:examples:uni}.
We adopt a randomized approach to   
the selection of the projection vectors $\mathbf{t}_1,\ldots,\mathbf{t}_\tau$, avoiding curse of dimensionality issues as the dimension $d$ increases (\citeNP{Nolan08}).

\section{Evaluation of model and sampler performance}
\label{sec:examples}

We now analyze the performance of the Bayesian $\alpha$-stable models and likelihood-free sampler in a sequence of simulation studies.
For the univariate model, we evaluate the capability of the summary statistics $S_1,\ldots,S_5$, and contrast the results with the samplers of \shortciteN{buckle95}  and \shortciteN{lombardi07}. The performance of the multivariate model under the statistics $S_6$ is then considered for two and three dimensions.

In the following, we implement the likelihood-free sequential Monte Carlo algorithm of \shortciteN{peters+fs08} (Appendix A) in order to simulate from the likelihood-free approximation to the true posterior $\pi_{LF}(\theta|y)\approx\pi(\theta|y)$ given by  (\ref{eqn:marginal-posterior}). This algorithm samples directly from $\pi_{LF}(\theta|y)$ using density estimates based on $P\geq 1$ Monte Carlo draws $x^1,\ldots,x^P\sim\pi(x|\theta)$ from the model.
We define $\pi_\epsilon(y|x,\theta)$ as a Gaussian 
kernel so that the 
summary statistics $S(y)\sim N(S(x),\epsilon^2\Sigma)$ for a suitably chosen $\Sigma$. 
All inferences are based on $N=1000$ particles drawn from $\pi_{LF}(\theta|y)$.
Detail of algorithm implementation is removed to Appendix A for clarity of exposition.

\subsection{Univariate summary statistics and samplers}
\label{sec:examples:uni}

We simulate $n=200$ observations, $y$, from a univariate $\alpha$-stable distribution with parameter values $\alpha=1.7$, $\beta=0.9$, $\gamma=10$ and $\delta=10$. We then implement  the likelihood-free sampler targeting $\pi_{LF}(\theta|y)$ for each of the univariate summary statistics 
$S_1$-$S_5$ described in Section \ref{sec:uni-ss},
with uniform priors for all parameters (Section \ref{sec:uni}). Alternative prior specifications were investigated  (\citeNP{lombardi07}; \citeNP{Nolan97}), with little impact on the results.

Posterior minimum mean squared error (MMSE) estimates for each parameter, averaged over $10$ sampler replicates
are detailed in Table \ref{tab1}.
Monte Carlo standard errors are reported in parentheses.  
The results indicate that all summary vectors apart from $S_5$ estimate $\alpha$ and $\delta$ parameters well,
and
for $\gamma$, $S_3$ and $S_4$ perform poorly. Only $S_1$ gives reasonable results for $\beta$ and for all parameters jointly.
Figure \ref{fig3} illustrates a progression of the MMSE estimates of each parameter using $S_1$, from the
likelihood-free SMC sampler output for each sampler replicate.
As the sampler progresses, the scale parameter
$\epsilon$ decreases, and the MMSE estimates
identify the true parameter values as the likelihood-free posterior approximation improves.

The results in Table \ref{tab1} are based on using $P=1$ Monte Carlo draws from the model to estimate $\pi_{LF}(\theta|y)$ (c.f. \ref{eqn:marginal-posterior}) within the likelihood-free sampler. Repeating the above study using $P\in\{5, 10, 20\}$ produced very similar results, and so we adopt $P=1$ for the sequel as the most computationally efficient choice.

For comparison, we also implement the auxiliary variable Gibbs
sampler of \shortciteN{buckle95} and the MCMC inversion and series
expansion sampler of \shortciteN{lombardi07}, based on
chains of length 100,000 iterations (10,000 iterations burnin), and using their respective prior specifications.
The Gibbs sampler performed poorly for most parameters.
The MCMC  method
performed better, but has larger standard errors than the likelihood-free sampler using $S_1$.

The MCMC sampler \shortcite{lombardi07} 
performs likelihood evaluations via inverse Fast Fourier transform 
(FFT) with approximate tail evaluation 
using Bergstrom 
expansions. This approach is sensitive to $\alpha$, which determines
the
threshold between the FFT and
the series expansion.
Further, as the tail becomes fatter, a finer spacing of FFT abscissae is required to
 control the bias introduced outside of the
Bergstrom series expansion,
significantly increasing computation. 
Overall, this sampler worked reasonably for $\alpha$ close to 2, though with deteriorating performance as $\alpha$ decreased.
The Gibbs sampler 
\shortcite{buckle95} performed extremely poorly for most settings and datasets, even when using their
proposed change of variables transformations. 
As such, the results in Table \ref{tab1} represent simulations under which both Gibbs and MCMC samplers performed credibly, thereby typifying their best case scenario performance.

\subsection{Multivariate samplers}
\label{sec:ss;mv}

We consider varying numbers of discrete spectral
masses, $k$, 
in the approximation to the spectral measure
$
\Gamma\left(  \cdot\right)  =\sum\limits_{j=1}^{k}w_{j}\delta_{\mathbf{s}_{j}
}(\cdot).
$
We assume that the number of spectral masses 
is known \textit{a priori}, and denote the $d$-variate $\alpha$-stable
distribution 
by
$\bm{S}_{\alpha}\left(d,k,\boldsymbol{w},\boldsymbol{\phi}_{1:k},\boldsymbol{\mu}^0\right)$. 
Priors are specified in Section \ref{sec:mvas}. Following the analysis of Section \ref{sec:examples:uni}, we incorporate
 $S_1$ within the
 summary vector $S_6$.


For
datasets of size 
$n=200$, we initially consider the performance of the bivariate $\alpha$-stable model, $\bm{S}_{\alpha}\left(2,k,\boldsymbol{w},\boldsymbol{\phi}_{1:k},\bm{0}\right)$, for $k=2$ and $3$ spectral masses, with respect to parameter estimation and the impact of the number of projection vectors $\mathbf{t}_1,\ldots,\mathbf{t}_\tau$.

The true and mean MMSE estimates of 
each parameter, placing projection vectors at the true locations of the spectral masses,
are presented  in Table \ref{tab2}. In addition, 
results are detailed using  $\tau=2, 5, 10$ and $20$ randomly (uniformly) placed projection vectors, in order to evaluate the impact of spectral mass location uncertainty.
The likelihood-free sampler output results in good MMSE parameter estimates, 
even for 2 randomly placed projection vectors. The parameter least accurately estimated is the 
location vector, $\boldsymbol{\mu}^0$. Directly summarized by a sample mean in $S_6(\cdot)$, estimation of location requires a large number of observations when the data have heavy tails.

Figure \ref{fig6} illustrates progressive sampler performance for the $\bm{S}_{\alpha}\left(2,2,\boldsymbol{w},\boldsymbol{\phi}_{1:2},\bm{0}\right)$ model, with $\alpha=1.7$, $\boldsymbol{w}=(\pi/4,\pi)$ and  $\boldsymbol{\phi}_{1:2}=(\pi/4,\pi)$.
Each circular scatter plot presents MMSE estimates of weight (radius) and angles (angle) of the two spectral masses, based on 10 sampler replicates. The sequence of plots (a)--(d) illustrates the progression of the parameter estimates as the scale parameter $\epsilon$ (of $\pi_\epsilon(y|x,\theta)$) decreases (and hence the accuracy of the likelihood-free approximation $\pi_{LF}(\theta|y)\approx\pi(\theta|y)$ improves).
As $\epsilon$ decreases, there is a clear movement
of the MMSE estimates towards the true angles and weights,
indicating appropriate sampler performance.


With simulated datasets of size $n=400$, we extend the previous bivariate study to 3 dimensions, with $k=2$ discrete spectral masses.
The true parameter values, and posterior mean MMSE estimates and associated standard errors, based on 10 sampler replicates, are presented in 
Table \ref{tab3}.
Again, reasonable parameter estimates are obtained (given finite data), with location ($\boldsymbol{\mu}^0$) again the most difficult to estimate.

In analogy with Figure \ref{fig6}, progressive sample performance for the first spectral mass (with $w_1=0.7$ and $\boldsymbol{\phi}_1=(\pi/4,\pi))$ for decreasing scale parameter $\epsilon$ is shown in Figure \ref{fig7}.
Based on 200 replicate MMSE estimates (for visualization purposes), the shading of each point indicates the value of $w_1$ as a percentage (black=0\%, white=100\%), and the location on the sphere represents the angles $\boldsymbol{\phi}_1$.
For large $\epsilon$, the MMSE estimates for location are
uniformly distributed over the sphere, and the associated weight takes the full range of possible values, $0-100$\%.
As $\epsilon$ decreases, the estimates of spectral mass location and weight become strongly localized and centered on the true parameter values,
again indicating appropriate sampler performance. Similar images are produced for the second discrete spectral mass.

\section{Analysis of exchange rate daily returns}
\label{sec:realdata}

Our data consist of daily exchange rates for 5 different currencies
recorded in GBP between 1 January 2005 and 1 December 2007. 
The data involve 1065 daily-averaged LIBOR (London interbank offered rate) observations $y'_1,\ldots,y'_{1065}$. 
The standard  log-transform 
generates
a log returns series 
$y_{t}=\ln\left( y'_{t+1}/y'_{t}\right)$.
Cursory examination of each returns series 
reveals clear non-Gaussian tails and/or skewness (Table \ref{tab:CurrencyExchange}, bottom}).

We initially model each currency series as independent draws from a univariate $\alpha$-stable distribution.
Posterior MMSE parameter estimates for each currency are given in Table \ref{tab:CurrencyExchange}, based on 10 replicate likelihood-free samplers using the $S_1$ summary vector.
For comparison, we also compute McCulloch's 
sample quantile based estimates
(derived from $S_1$, c.f. Section \ref{sec:uni-ss}), and
maximum likelihood estimates using J. P. Nolan's {\it STABLE} program (available online),
using the direct search SPDF option with search domains given by $\alpha\in(0.4,2]$, $\beta\in[-1,1]$, $\gamma
\in[0.00001,1]$ and  $\delta\in[-1,1]$.
Overall, there is good agreement between Bayesian, likelihood- and sample-based estimators. 
All currency returns distributions are significantly different
from Gaussian ($\alpha=2,\beta=0$), 
and
exhibit similar family parameter $(\alpha)$
estimates over this time period.
However, the GBP to
YEN conversion demonstrates a significantly asymmetry ($\beta$)
compared to the other currencies.

An interesting difference between the methods of estimation, is that McCulloch's estimates of $\alpha$ differ considerably from the posterior MMSE estimates, even though the latter are constructed using McCulloch's estimates directly as summary statistics, $S_1$.
One reason that the Bayesian estimates are more in line with the MLE's, is that likelihood-free methods largely ignore bias in estimators used as summary statistics (comparing the closeness between biased or unbiased estimators
will produce similar results -- consider comparing sample and maximum likelihood estimators of variance).


The multivariate
$\alpha$-stable distribution assumes that its marginal distributions, which are also $\alpha$-stable,
possess identical shape parameters.
This property implies important practical limitations, one of which is that
it is only sensible to jointly model data with similar marginal shape parameters. 
Accordingly, based on Table \ref{tab:CurrencyExchange},
we now consider a bivariate analysis of AUD and EURO currencies.
Restricting the analysis to the bivariate setting also permits comparison with 
the bivariate frequentist approach described in \shortciteN{Nolan97} using the \textit{MVSTABLE} software (available online).

A summary illustration of the discrete approximations to the underlying continuous spectral mass is shown in Figure \ref{fig9}. 
Assuming $k=3$ discrete spectral masses and based on 10 likelihood-free sampler replicates, the mean MMSE posterior estimates (solid black line) with mean $3\sigma$ posterior credibility intervals (dotted line), identify regions of high spectral mass located at 
2.7, 3.9 and 5.6, with respective weights 0.45, 0.2 and 0.35.
Broken lines in Figure \ref{fig9} denote the frequentist estimates of \shortciteN{Nolan97}, based on the identification of mass over an exhaustive mesh grid
using 40 (dashed line) and 80 (dash-dot line) prespecified grid locations (projections).

Overall, both approaches produce comparable summary estimates of the spectral mass approximation, although the likelihood-free models generate full posterior distributions, compared to Nolan's 
frequentist 
estimates. The assumption of $k=3$ discrete spectral masses provides a parsimonious representation of the actual spectral mass. For example, the spectral mass located at 
2.7
accounts for the first two/three masses based on Nolan's estimates (80/40 projections).
While the frequentist approach is computationally restricted to bivariate inference, the likelihood-free approach may naturally be applied in much higher dimensions.

\section{Discussion}

Statistical inference for $\alpha$-stable models is challenging due to the computational intractability of the density function. In practice this limits the range of models fitted, to  univariate and bivariate cases.
By adopting likelihood-free Bayesian methods we are able to circumvent this difficulty, and provide approximate, but credible posterior inference in the general multivariate case, at a moderate computational cost.
Critical to this approach is the availability of informative summary statistics for the parameters. We have shown  that multivariate projections of data onto the unit hypersphere, in combination with sample quantile estimators, are adequate for this task.

Overall, our approach has a number of advantages over existing methods. There is far greater sampler consistency than  alternative samplers, such as the auxiliary Gibbs or MCMC inversion plus series expansion samplers (\citeNP{buckle95}; \shortciteNP{lombardi07}).
It is largely independent of the complexities of the various parameterizations of the $\alpha$-stable characteristic function.
The likelihood-free approach is conceptually straightforward, and scales simply and is easily implemented in higher dimensions (at a higher computational cost).
Lastly, by permitting a full Bayesian multivariate analysis, the component locations and weights of a discrete approximation to the underlying continuous spectral density are allowed to identify those regions with highest posterior density in  a parsimonious manner. This is a considerable advantage over highly computational frequentist approaches, which require explicit calculation of the spectral mass over a deterministic and exhaustive grid (e.g. \shortciteNP{Nolan97}).

Each analysis in this article used many millions of data-generations from the model.
While computation for likelihood-free methods increases with model dimension and desired accuracy of the model approximation (through $\epsilon$), much of this 
can be offset through parallelization of the likelihood-free sampler \shortcite{peters+fs08}.

Finally, while we have largely focused on fitting $\alpha$-stable models in the likelihood-free framework, extensions to model selection through Bayes factors or model averaging are immediate. One obvious candidate in this setting is the unknown number of discrete spectral masses, $k$, in the approximation to the continuous spectral density.

\subsection*{Acknowledgments}

YF and SAS are supported by the Australian Research Council
Discovery Project scheme (DP0877432 \& DP1092805).
GWP is supported by an APA scholarship
and by the School of Mathematics and
Statistics, UNSW and CSIRO CMIS. We thank 
M. Lombardi for generously providing the use of his code \shortcite{lombardi07}, and
 M. Briers, S. Godsill, Xiaolin Lou, P.
Shevchenko and R. Wolpert, for thoughtful discussions.

\bibliographystyle{chicago}
\bibliography{a-stable}


\newpage

\subsection*{Appendix A}

\noindent{\bf SMC sampler PRC-ABC algorithm \shortcite{peters+fs08}} 
\vskip 0.3cm
\hrule width 430pt depth 1pt
\vskip 0.4cm
\begin{description}
\item[Initialization:] 
Set $t=1$ and specify tolerance schedule $\epsilon_1,\ldots,\epsilon_T$.\\
For $i=1,\ldots, N$, sample $\theta^{(i)}_1 \sim \pi(\theta)$,  and set
weights 
$W_1(\theta^{(i)}_1) = \pi_{LF,1}(\theta_1^{(i)}|y)/\pi(\theta^{(i)}_1)$.

\item[Resample:] Resample $N$ particles with respect to $W_t(\theta_t^{(i)})$ and set $W_t(\theta_t^{(i)})=\frac{1}{N},\\ i=1,\ldots,N$.

\item[Mutation and correction:] Set $t=t+1$ and $ i=1$:\\
\begin{tabular}{ll}
(a)& Sample  $\theta_t^{(i)}\sim M_t(\theta_t)$ and set weight for 
$\theta_t^{(i)}$ to  \\
&  \quad\quad $W_t(\theta_t^{(i)})=\pi_{LF,t}(\theta_t^{(i)} | y) /M_t(\theta_t^{(i)})$.\\
(b)& With probability $1-p^{(i)}=1-\min\{ 1,W_t(\theta_t^{(i)})/c_t\}$, reject $\theta_t^{(i)}$ and go to (a). \\
(c)& Otherwise, accept  $\theta_t^{(i)}$ and set  $W_t(\theta_t^{(i)}) =  W_t(\theta_t^{(i)})/p^{(i)}$.\\
(d)&    Increment $i=i+1$.   If $i\leq N$, go to (a).\\
(e)&  If  $t<T$ then go to Resample.
\end{tabular}
\end{description}
\vskip 0.0cm
\hrule width 430pt depth 1pt
\vskip 0.5cm

\noindent This algorithm samples $N$ weighted {\it particles} from a sequence of distributions $\pi_{LF,t}(\theta|y)$ given by (\ref{eqn:marginal-posterior}), where $t$ indexes a sequence of scale parameters $\epsilon_1\geq\ldots\geq\epsilon_T$. The final particles $\{(W_T(\theta^{(i)}_T),\theta_T^{(i)}):i=1,\ldots,N\}$, form a weighted sample from the target $\pi_{LF,T}(\theta|y)$ (e.g. \shortciteNP{peters+fs08}). The densities $\pi_{LF,t}(\theta|y)$ are estimated through the Monte Carlo estimate of the expectation (\ref{eqn:marginal-posterior}) based on $P$ draws $x^1,\ldots,x^P\sim\pi(x|\theta)$.

For the simulations presented we implement the following specifications: 
for univariate $\alpha$-stable models $\theta=(\alpha,\beta,\gamma,\delta)$ and for multivariate models
$\theta=(\boldsymbol{w},\boldsymbol{\phi}_{1:k},\boldsymbol{\mu}^0,\alpha)$;
we use $N=1000$ particles, initialized with samples from the prior; the function $\pi_\epsilon(y|x,\theta)$ is defined by $S(y)\sim N(S(x), \epsilon^2\hat{\Sigma})$ where $\hat{\Sigma}$ is an estimate of $\mbox{Cov}(S(x)|\hat{\theta})$ based on 1000 draws $x^1,\ldots,x^{1000}\sim\pi(x|\hat{\theta})$ given an approximate maximum likelihood estimate $\hat{\theta}$ of $\theta$ \cite{jiang+t04};
the mutation kernel
$
M_t(\theta_t) = \sum_{i=1}^N W^{(i)}_{t-1}(\theta^{(i)}_{t-1})\phi(\theta_t;\,\theta^{(i)}_{t-1},\Lambda)
$
is a density estimate of the previous particle population $\{(W_{t-1}(\theta^{(i)}_{t-1}),\theta_{t-1}^{(i)}):i=1,\ldots,N\}$,
with a Gaussian kernel density $\phi$ with covariance $\Lambda$;
for univariate $\alpha$-stable models $\Lambda=\mbox{diag}(0.25,0.25,1,1)$, and for multivariate models $\Lambda=\mbox{diag}(1,\ldots,1,0.25)$ (with Dirichlet proposals and kernel density substituted for $\boldsymbol{w}$);
the sampler particle rejection threshold is adaptively determined as the  $90^{th}$ quantile of the weights
$c_t =
\widehat{q}_{0.9}(\{W^{(i)}_t(\theta_t^{(i)})\}),
$
where $\{W^{(i)}_t(\theta_t^{(i)})\}$ are the $N$ particle weights {\it prior} to particle rejection (steps (b) and (c)) at each sampler stage $t$ (see \shortciteNP{peters+fs08}).

For each analysis we implement 10 independent samplers (in order to monitor algorithm performance and Monte Carlo variability), each with the deterministic scale parameter sequence:
$\epsilon_t\in
\{1000,900,\ldots,200,100,99,\ldots,11,10,9.5,9,\ldots,5.5,5,4.95,$ $\ldots,3.05,3,2.99,2.98,\ldots,0.01, 0\}$. 
However, we adaptively terminate all samplers at the largest $\epsilon$ value such that the effective sample size (estimated by $[\sum_{i=1}^N [W^{(i)}_{t}(\theta_{t}^{(i)})]^2]^{-1}$) consistently drops below $0.2N$ over all replicate sampler implementations.

\subsection*{Appendix B: Data generation}

\noindent \textit{Simulation of univariate $\alpha$-stable data (\shortciteNP{dumouchel75}, \shortciteNP{chambers+ms76}) }

\begin{enumerate}
\item Sample $W \sim \mbox{Exp}(1)$ to obtain $w$

\item Sample $U \sim \mbox{Uniform}[-\pi/2,\pi/2]$ to obtain $u$

\item Apply transformation to obtain sample $\overline{y}$
\[
\overline{y}=\left\{
\begin{array}
[c]{cll}
&S_{\alpha,\beta}\frac{\sin\alpha\left(
u+B_{\alpha,\beta}\right) }{(\cos u)^{\alpha/2}}\left[
\frac{\cos\left(  u-\alpha\left( u+B_{\alpha,\beta}\right)
\right)  }{w}\right]  ^{\frac{1-\alpha
}{\alpha}}&\text{ \ \ if }\alpha\neq1\\
&\frac{2}{\pi}\left[  \left(  \frac{\pi}{2}+\beta u\right) \tan
u-\beta\ln\frac{\frac{\pi}{2}w_{i}\cos u}{\frac{\pi}{2}+\beta u
}\right] & \text{ \ \ if }\alpha=1
\end{array}
\right.
\]
with $S_{\alpha,\beta}=\left(  1+\beta^{2}\tan^{2}\left(  \frac{\pi\alpha}
{2}\right)  \right)  ^{-1/2\alpha}$ and
$B_{\alpha,\beta}=\frac {1}{\alpha}\arctan\left(  \beta\tan\left(
\frac{\pi\alpha}{2}\right) \right) .$ In this case
$\overline{y}$ will have distribution defined by
$\Phi_{X}\left(  t\right)  $ with parameters
$\left(  \alpha,\beta,1,0\right)  $.

\item Apply transformation to obtain sample
$y=\gamma\overline{y}+\delta$ with parameters
$\left(  \alpha,\beta,\gamma,\delta\right)  $.
\end{enumerate}

\vskip 0.5cm

\noindent \textit{Simulation of $d$-dimensional, multivariate $\alpha$-stable data  (\shortciteNP{nolan07})}

\begin{enumerate}
\item Generate $Z_{1},...,Z_{k}$ i.i.d. random variables from the univariate
$\alpha$-stable distribution with parameters $(\alpha,\beta,\gamma,\delta)=(\alpha, 1, 1, 0)$.

\item Apply the transformation
\[
Y\mathbf{=}\left\{
\begin{array}{ll}
\sum\limits_{j=1}^{k}w_{j}^{1/\alpha}Z_{j}\mathbf{s}_{j}+\boldsymbol{\mu}^{0}\text{
\ \ } & \alpha\neq1\\
\sum\limits_{j=1}^{k}w_{j}\left(  Z_{j}+\frac{2}{\pi}\ln\left(
w _{j}\right)  \right)  \mathbf{s}_{j}+\boldsymbol{\mu}^{0}\text{ \ \
} & \alpha=1
\end{array}
\right.  
\]
\end{enumerate}
with $\boldsymbol{s}_1,\ldots,\boldsymbol{s}_k,\boldsymbol{\mu}^0\in\mathbb{S}^d$.
Note that while the complexity for generating realizations from a multivariate $\alpha$-stable distribution is
linear in the number of point masses ($k$) in the spectral
representation per realization, this method is strictly only exact for discrete spectral measures.\\
\\

\begin{table}[ptbh]
\begin{center}
{\footnotesize {\scriptsize {\
\begin{tabular}{|c|c|c|c|c|c|c|c|}
\hline
 & {Buckle} & {Lombardi} & $S_1$ & $S_2$ & $S_3$ & $S_4$ & $S_5$\\ 
\hline
$\alpha$ (1.7)  & 1.77 (0.18) & 1.62 (0.10) & 1.69 (0.06) & 1.65 (0.07) &1.70 (0.06) &1.71 (0.04) & 1.56 (0.05)\\
$\beta$ (0.9)  &0.54 (0.21) & 0.86 (0.18) & 0.86 (0.10) & 0.65 (0.13) &0.31 (0.09) & 0.38 (0.12) & 0.49 (0.11)\\
$\gamma$ (10.0) &18.17 (6.19) & 9.59 (2.16) & 9.79 (0.21) &10.44(0.56) & 38.89 (6.34) & 39.12 (5.92) & 9.34 (0.14)\\
$\delta$ (10.0) &12.30 (4.12) & 9.70 (2.19) &10.64 (0.83) &9.31 (0.86) &10.25 (0.98) & 10.83 (1.34) & 11.18 (1.05)\\
\hline
\end{tabular}
 } }  }
\end{center}
\caption{\small Means and standard errors (in parentheses) of posterior MMSE estimates of $\alpha$, $\beta$, $\gamma$ and $\delta$ under the univariate $\alpha$-stable model,
based on 10 sampler replicates.
Parameter values used for data simulation are given in the left column. Comparisons are between the auxiliary variable Gibbs sampler method of 
Buckle (1995), the inversion MCMC method of Lombardi (2007), and the 
likelihood-free method, using summary statistics $S_1$--$S_5$.
}\label{tab1}
\end{table}
}

{\footnotesize
\begin{table}[ptbh]
\begin{center}
{\footnotesize {\scriptsize {\
\begin{tabular}{|ccccccccc|}
\hline  & $\alpha$ & $\mu_1^{0}$ &
$\phi^{1}$ & $\phi^{2}$ & $\phi^{3}$ &$w_{1}$ & $w_{2}$ & $w_{3}$ \\
True:  $k=2$ & 1.7 & 0 & $\frac{\pi}{4}$ & $\pi$ & -- & 0.6 & 0.4 & --\\
True:  $k=3$ & 1.7 & 0 & $\frac{\pi}{4}$ & $\pi$ & $\frac{3\pi}{2}$ & 0.3 & 0.25 & 0.45\\
\hline $k$  & $\widehat{\alpha}$ &
$\widehat{\mu_1^{0}}$ & $\widehat{\phi^{1}}$ &
$\widehat{\phi^{2}}$ & $\widehat{\phi^{3}}$ &
$\widehat{w_{1}}$ & $\widehat{w_{2}}$ & $\widehat{w_{3}}$ \\
\hline
\multicolumn{9}{|c|}{Projection vectors at locations of true spectral masses.}\\
2  &  1.66 (0.04) &  0.16 (0.19) & 0.81 (0.65) & 3.19 (0.46) & -- &  0.55 (0.06) &  0.45 (0.05) & --\\
3  &  1.79 (0.02) &  0.36 (0.18) & 0.84 (0.27) & 3.18 (0.29) & 4.91 (0.24) & 0.35 (0.05) & 0.25 (0.04) & 0.40 (0.05)\\
\hline
\multicolumn{9}{|c|}{2 projection vectors}\\
2  &  1.67 (0.06) &  -0.13 (0.16) & 0.73 (0.55) & 3.58 (0.57) & -- & 0.58 (0.09) & 0.42 (0.10) & --\\
3 &  1.76 (0.05) &  -0.16 (0.26) & 0.91 (0.66) & 3.65 (0.62) & 4.85 (0.55) & 0.36 (0.10) & 0.24 (0.09) & 0.40 (0.08)\\
\hline
\multicolumn{9}{|c|}{5 projection vectors}\\
2  &  1.71 (0.05) &  0.08 (0.14) & 0.71 (0.60) &  3.80 (0.67) & -- & 0.60 (0.07) & 0.40 (0.09) & --\\
3  &  1.75 (0.04) &  0.29 (0.17) & 0.86 (0.62) &  3.68 (0.52) & 4.82 (0.41) & 0.35 (0.09) & 0.20 (0.07) & 0.42 (0.09)\\
\hline
\multicolumn{9}{|c|}{10 projection vectors}\\
2 &  1.72 (0.02) &  0.21 (0.21) & 0.75 (0.32) & 3.31 (0.32) & -- & 0.59 (0.05) & 0.41 (0.07) & --\\
3  &  1.73 (0.03) &  0.25 (0.16) & 0.76 (0.44) & 3.31 (0.48) & 4.78 (0.19) & 0.34 (0.07) & 0.24 (0.04) & 0.42 (0.05)\\
\hline
\multicolumn{9}{|c|}{20 projection vectors}\\
2  &  1.71 (0.03) & -0.14 (0.14) & 0.76 (0.36) & 3.21 (0.23) & -- & 0.63 (0.03) & 0.37 (0.03) & --\\
3  &  1.72 (0.02) & 0.18 (0.23) &  0.77 (0.32) & 3.25 (0.31) & 4.75 (0.15) & 0.34 (0.04) & 0.23 (0.03) & 0.43 (0.03)\\
\hline
\end{tabular} } }  }
\end{center}
\caption{Mean MMSE parameter estimates (and standard errors) for the bivariate $\alpha$-stable
$\bm{S}_{\alpha}\left(2,k,\boldsymbol{w},\boldsymbol{\phi}_{1:k},\boldsymbol{\mu}^0\right)$ model, for $k=2,3$ discrete spectral masses,
calculated over $10$
replicate samplers. Projections vectors are placed at the true, and 2, 5, 10 and 20 randomly selected spectral mass locations. The true value of $\boldsymbol{\mu}^0$ is the origin.} \label{tab2}
\end{table}
}

{\footnotesize
\begin{table}[ptbh]
\begin{center}
{\footnotesize {\scriptsize {\
\begin{tabular}
{|ccccccccc|}
\hline  & $\alpha$ & $\mu_1^{0}$ & $\phi_1^{1}$ & $\phi_2^{1}$ & $\phi_1^{2}$ & $\phi^{2}_{2}$ & $w_{1}$ & $w_{2}$ \\
True:  $k=2$ & 1.7 & 0 & $\frac{\pi}{4}$ & $\pi$ & $\frac{\pi}{2}$ & $\frac{3\pi}{2}$ & 0.3 & 0.7\\
\hline $k$  & $\widehat{\alpha}$ &
$\widehat{\mu_1^{0}}$ & $\widehat{\phi_{1}^{1}}$ &
$\widehat{\phi_{2}^1}$ & $\widehat{\phi_{1}^2}$ &
$\widehat{\phi_2^2}$ & $\widehat{w_{1}}$ & $\widehat{w_{2}}$ \\
\hline
\multicolumn{9}{|c|}{20 projection vectors}\\
2 & 1.71 (0.02) & 0.53 (0.89) & 1.12 (0.34) & 3.81 (0.45) & 1.84 (0.54) & 4.24 (0.69) & 0.28 (0.06) & 0.72 (0.05)\\
\hline
\end{tabular} } }  }
\end{center}
\caption{ Mean MMSE parameter estimates (and standard errors) for the trivariate $\alpha$-stable
$\bm{S}_{\alpha}\left(3,2,\boldsymbol{w},\boldsymbol{\phi}_{1:2},\boldsymbol{\mu}^0\right)$ model, with
$k = 2$ discrete spectral masses, calculated over $10$ replicate samplers.
The true value of $\boldsymbol{\mu}^0$ is the origin.
} \label{tab3}
\end{table}
}

{\footnotesize
\begin{table}[ptbh]
\begin{center}
{\footnotesize {\scriptsize {\
\begin{tabular}
[c]{|cc|ccccc|}\hline & \multicolumn{1}{c}{} &
\multicolumn{5}{|c|}{{ Currency Exchange from GBP
to}}\\\cline{3-7}  & \multicolumn{1}{c|}{} & { AUD} &
\multicolumn{1}{|c}{{ CNY}} & \multicolumn{1}{|c}{{ EURO}} &
\multicolumn{1}{|c}{{ YEN}} & \multicolumn{1}{|c|}{{
USD}}\\
\hline\hline
 & $\widehat{\alpha}$ & 1.56 (0.03) & 1.57 (0.02) & 1.62 (0.04) & 1.51 (0.04) & 1.53 (0.02) \\
Likelihood & $\widehat{\beta}$ & 0.06 (0.03) & 0.01 (0.009) & -0.007 (0.08) & -0.26 (0.09) & -0.04 (0.03) \\
free & $\widehat{\gamma}$ & 0.004 (4e-4) & 0.003 (2e-4) & 0.004 (1e-4) & 0.003 (1e-4) & 0.004 (3e-4) \\
& $\widehat{\delta}$ & 0.02 (0.01) & 0.001 (0.0006) & -0.03 (0.09) & -0.06 (0.08) & -0.02 (0.07) \\
\hline\hline
& $\widehat{\alpha}$ & 1.61 (0.05) & 1.50 (0.05) & 1.65 (0.05) & 1.66 (0.04) & 1.57 (0.05)\\
{ MLE} & $\widehat{\beta}$ & 0.08 (0.11) & -0.01 (0.10) & -0.10 (0.12) & -0.46 (0.11) & -0.01 (0.11)\\
& $\widehat{\gamma}$ & 0.002 (7e-5) & 0.002 (6e-5) & 0.001 (4e-5) & 0.002 (4e-5) & 0.002 (1e-4)\\
& $\widehat{\delta}$ & -2e-4 (1e-4)  & -2e-5 (1e-4) & 8e-5 (7e-5) & 6e-4 (1e-4) & 5e-5 (9e-5)\\
\hline\hline
McCulloch's  & $\widehat{\alpha}$ & 1.39 & 1.38 & 1.47 & 1.38 & 1.39 \\
quantile & $\widehat{\beta}$ & 0.08 & -0.003 & -0.04 & -0.18 & 0.001 \\
estimates & $\widehat{\gamma}$ & 0.002 & 0.002 & 0.001 & 0.002 & 0.002\\
 & $\widehat{\delta}$ & -4e-5 & 1e-6 & 1e-5 & 2e-4 & 5e-7\\\hline\hline
{Kurtosis} &  & { 8.39} & { 9.11} & { 15.60}& { 6.29} & { 4.98}\\
{Skewness} &  & { 0.69} & { -0.42} &{ -0.03} & { -0.79} & { 0.11}\\
{ Std. dev.} &  & { 0.004} & { 0.004} & { 0.003} & {0.004} & { 0.003}\\
{ Mean} &  & { -4e-5} & { -4e-5} &{ -9e-6} & { 1e-4} & { 7e-5}\\
\hline
\end{tabular} } }  }
\end{center}
\caption{Posterior MMSE estimates (Monte Carlo errors) from the likelihood-free model, and maximum likelihood estimates (standard deviation). MLE's,  parameter estimates using McCulloch's quantile (McCulloch, 1998), and sample statistics  (mean, standard deviation, skewness and kurtosis) obtained from J. P. Nolan's {\it STABLE}
software,
available at {\tt academic2.american.edu/}$\sim${\tt jpnolan}.
}
\label{tab:CurrencyExchange}
\end{table}
}

\begin{center}
\begin{figure}[ptbh]
    \centering
 \includegraphics[width=\textwidth]{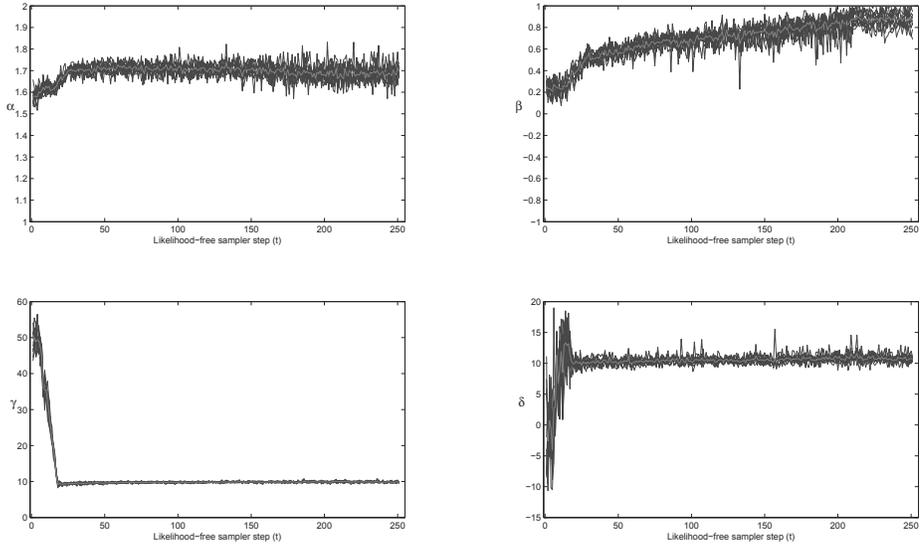}
        \caption{\small Traces of posterior MMSE estimates of $\alpha$, $\beta$, $\gamma$ and $\delta$ under the univariate $\alpha$-stable model and likelihood-free sampler (summary statistics $S_1$), based on 10 sampler replicates.
        Traces are shown as a function ($x$-axis) of sampler progression ($t$) and scale parameter reduction $\epsilon_{t}<\epsilon_{t-1}$. Parameter values used for data generation are $\alpha=1.7$, $\beta=0.9$, $\gamma=10$ and $\delta=10$.
        }
    \label{fig3}
\end{figure}
\end{center}

\begin{figure}[ptbh]
    \centering
        \includegraphics[width=.8\textheight]{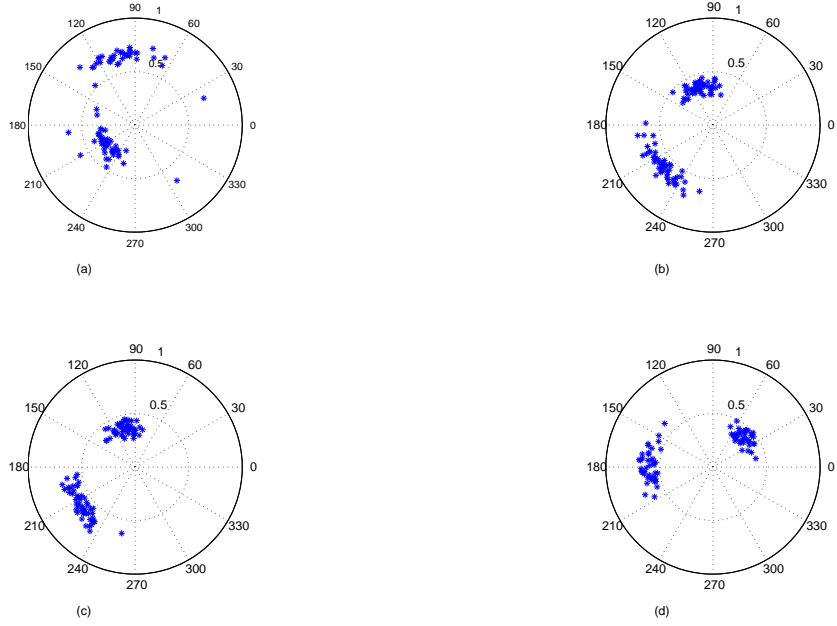}
        \caption{
        Circular scatter plot of MMSE estimates for $k=2$ spectral mass angles (angle) and weights (radius) for bivariate $\alpha$-stable $\bm{S}_{\alpha}\left(2,2,\boldsymbol{w},\boldsymbol{\phi}_{1:2},\boldsymbol{0}\right)$ model, with $\alpha=1.7$, $\boldsymbol{w}=(0.4,0.6)$ and $\boldsymbol{\phi}_{1:2}=(\pi/4,\pi)$.
        Plots (a)--(d) demonstrate evolution of the estimates for decreasing scale parameter values $\epsilon$, based on 10 sampler replicates.
}
    \label{fig6}
\end{figure}

\begin{figure}[ptbh]
    \centering
        \includegraphics[width=.6\textheight]{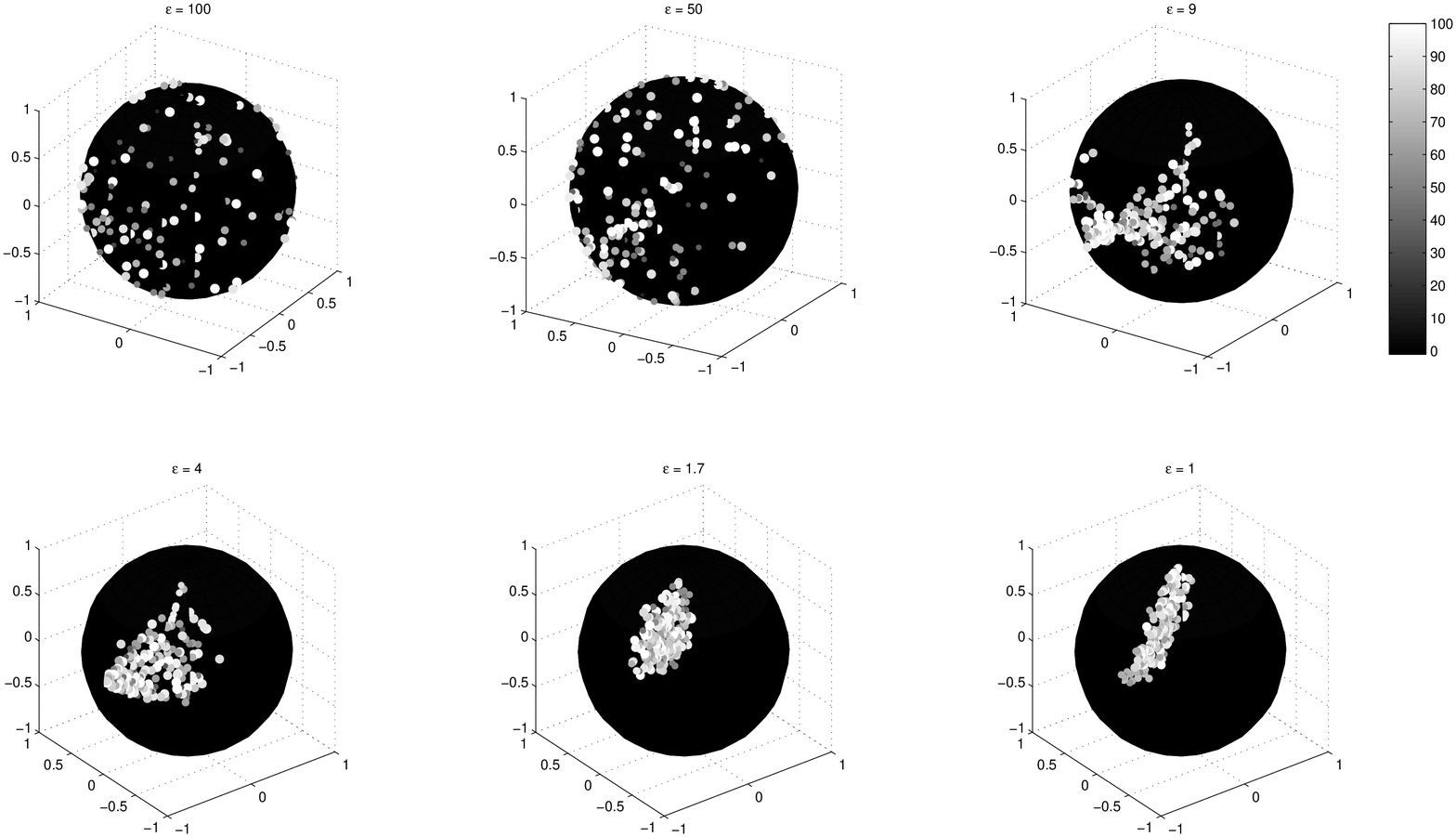}
        \caption{Spherical heat map of MMSE estimates for the first of $k=2$ discrete spectral masses, $\boldsymbol{\phi}_{1}$, for the trivariate $\alpha$-stable
        $\bm{S}_{\alpha}\left(3,2,\boldsymbol{w},\boldsymbol{\phi}_{1:2},\boldsymbol{0}\right)$ model. True values of the first spectral mass are $w_1=0.7$ (70\%) and $\boldsymbol{\phi}_1=(\pi/4,\pi)$. Point shading indicates MMSE value of $w_1$ as a percentage.
       The plots demonstrate
        the evolution of the estimates for decreasing scale parameter values $\epsilon$, based on 200 sampler replicates.
        }
    \label{fig7}
\end{figure}

\begin{figure}[ptbh]
    \centering
        \includegraphics[width=0.75\textheight]{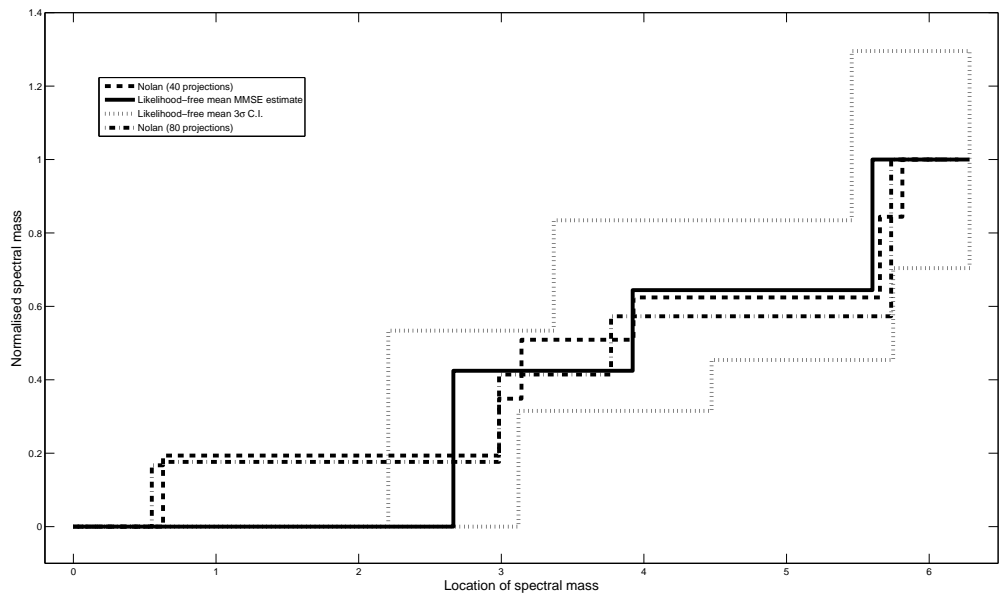}
        \caption{Estimates of spectral mass location (x-axis) and cumulative weight (y-axis) for AUD and EURO currencies data. Solid line denotes mean posterior MMSE estimates of likelihood-free SMC sampler output, and dotted line illustrates mean $3\sigma$ posterior credibility intervals, based on $k=3$ discrete spectral masses, 20 randomly placed projection vectors and 10 replicate samplers.
        Broken lines denote estimate of spectral mass using J. P. Nolan's {\it MVSTABLE} software, available at {\tt academic2.american.edu/}$\sim${\tt jpnolan}, with  (dashed line) 40 deterministic projection locations and (dash-dot line) 80 projection locations.
        }
    \label{fig9}
\end{figure}

\end{document}